\documentclass[10pt,onecolumn,twoside]{IEEEtran}
\ifCLASSINFOpdf
\else
   \usepackage[dvips]{graphicx}
\fi
\usepackage{url}

\usepackage{bbding}
\usepackage{graphicx}
\usepackage{times}
\usepackage{epsfig}
\usepackage{graphicx}
\usepackage{amsmath}
\usepackage{amssymb}
\usepackage{multirow}
\usepackage{booktabs}
\usepackage{algorithmic}
\usepackage{algorithm}
\usepackage{subfigure}
\usepackage{tablefootnote}
\usepackage{booktabs}
\usepackage{multirow}
\usepackage{cite}
\usepackage{bbding}
\usepackage{times}
\usepackage{multirow}
\usepackage{booktabs}

\usepackage{textcomp}
\usepackage{xcolor}

\begin{document}

\title{Digital-SC: Digital Semantic Communication with Adaptive Network Split and Learned Non-Linear Quantization\\}

\author{\IEEEauthorblockN{Lei Guo, Wei Chen, \IEEEmembership{Senior Member, IEEE}, Yuxuan Sun, \IEEEmembership{Member, IEEE}, Bo Ai, \IEEEmembership{Fellow, IEEE}}

\thanks{All authors are with School of Electronic and Information Engineering, Beijing Jiaotong University, Beijing, China.\\
Email: \{leiguo, weich, yxsun, boai\} @bjtu.edu.cn.}
}

\maketitle

\begin{abstract}

Semantic communication, an intelligent communication paradigm that aims to transmit useful information in the semantic domain, is facilitated by deep learning techniques. Robust semantic features can be learned and transmitted in an analog fashion, but it poses new challenges to hardware, protocol, and encryption. In this paper, we propose a digital semantic communication system, which consists of an encoding network deployed on a resource-limited device and a decoding network deployed at the edge. To acquire better semantic representation for digital transmission, a novel non-linear quantization module is proposed to efficiently quantize semantic features with trainable quantization levels. Additionally, structured pruning is incorporated to reduce the dimension of the transmitted features. We also introduce a semantic learning loss (SLL) function to reduce semantic error. To adapt to various channel conditions and inputs under constraints of communication and computing resources, a policy network is designed to adaptively choose the split point and the dimension of the transmitted semantic features. Experiments using the CIFAR-10 and ImageNet dataset for image classification are employed to evaluate the proposed digital semantic communication network, and ablation studies are conducted to assess the proposed quantization module, structured pruning, and SLL.
 
\end{abstract}


\IEEEpeerreviewmaketitle

\section{Introduction}
With the development of the fifth generation (5G) wireless communications, artificial intelligence (AI) and other supporting technologies, various mission-critical applications appear, such as autonomous driving, smart industry and tele-health \cite{10328187, 2018Semantic}. In the current communication systems, the raw data generated by these applications are transmitted from devices to edge or cloud computing servers, where advanced AI algorithms are used for further analysis. The traditional communication system aims at reproducing exactly or approximately the source data. However, with limited spectrum resources, it is extremely challenging to meet the stringent service requirements of these applications, such as massive connectivity, real-time data transmission and processing, etc.

Recently, semantic communication has attracted widespread research interest \cite{10328187,9955525,9852388,9814491,9653664,9830752,9796572,9771334}, which integrates advanced communication and AI technologies to extract and then transmit the desired \emph{meaning} of the raw data given certain tasks, e.g., classification and object detection. Therefore, semantic communications are also known as task-oriented communications \cite{9955525,9653664,9830752,9796572}. Unlike classical communications, semantic communications no longer aim to minimize the bit error rate. Instead, the main goal of semantic communications is to achieve the reliable transmission of useful information in the semantic domain. Accordingly, the required bandwidth can be significantly reduced, making semantic communications a promising solution to mission-critical applications.

\begin{figure}
\centering
\includegraphics[width=0.5\linewidth]{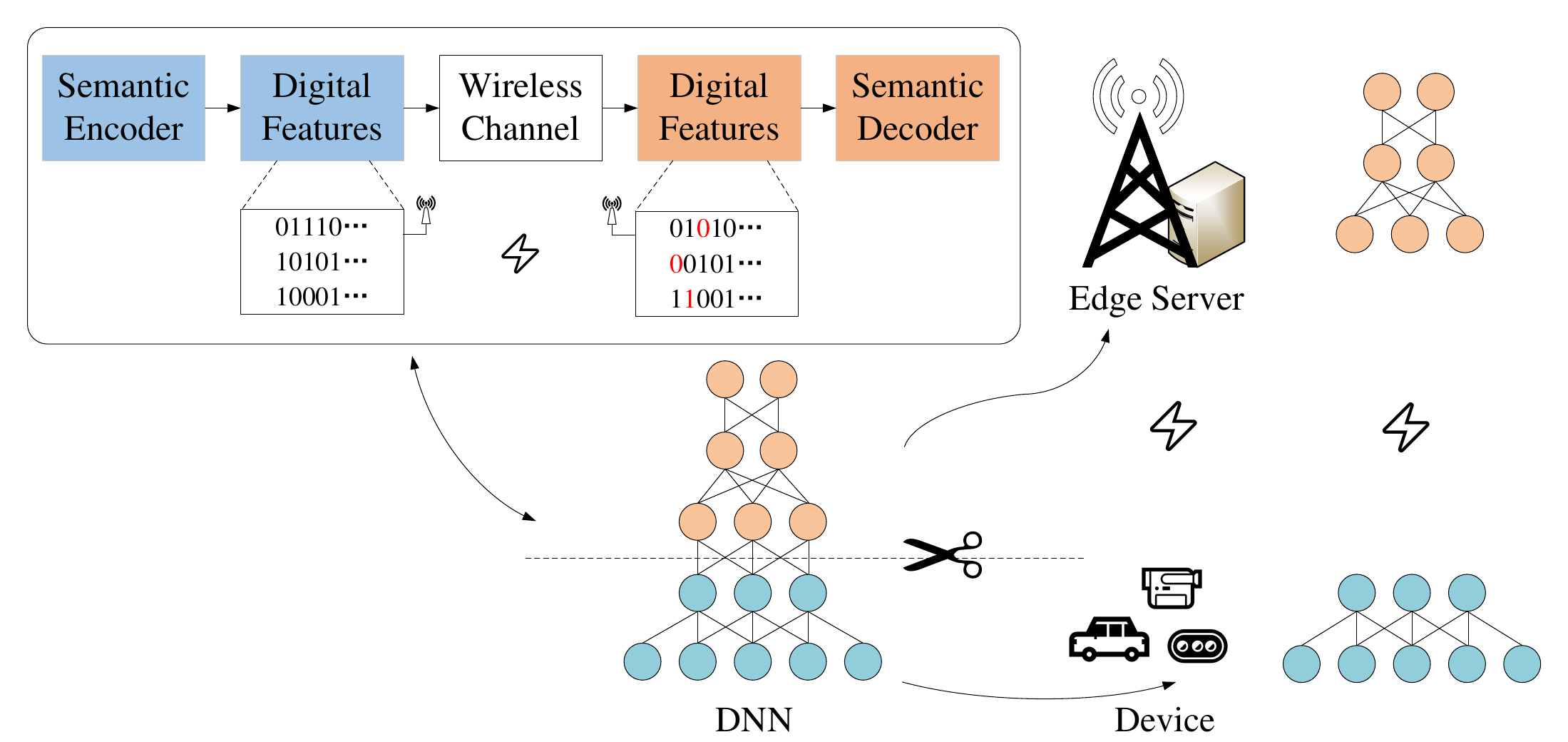}
\caption{The digital semantic communication system for device-edge cooperative inference. A deep neural network (DNN) is split into a device network and an edge network. The semantic features are quantized into bit sequences and transmitted by the digital wireless communication system.}
\label{fig:system}
\end{figure}

A typical scenario of semantic communications is the cooperative inference by devices and edge servers \cite{9653664,9830752}. As shown in Fig. \ref{fig:system}, the device with limited computing power extracts task-specific semantic features from the raw data using deep learning (DL) algorithms, and transmits them to the edge server via the wireless channels. Then, the edge server with powerful computing capability analyses the received semantic information to complete the inference task. Semantic communication systems have been proposed for different types of data sources, including image \cite{JankowskiMikolaj0Wireless,2020BottleNet,2019Improving}, text \cite{2021Deep,9955525,10118965} and speech signals \cite{2021Semantic,9953316}. In order to enhance its ability to adjust to various signal-to-noise ratios (SNR), the attention mechanism has been incorporated to assign weights to the features based on their significance and the channel conditions \cite{9438648}.

The studies discussed above consider analog communication schemes, where the feature vectors are directly mapped into analog symbols for transmission without being converted to bits at any stage \cite{JankowskiMikolaj0Wireless}. However, digital communication has many advantages over analog communication, including higher error correction capability, stronger anti-interference ability, higher security, etc. Digital schemes have been proposed in \cite{2018JALAD,2019BottleNet} for distributed inference, which use linear quantization functions \cite{2016DoReFa} and convert the full-precision features into bit sequences. Digital semantic communication has been further studied in \cite{XieHuiqiang2021A} for the Internet of Things (IoTs) by incorporating the linear quantization and deep neural network (DNN) compression. To improve the flexibility, a novel digital semantic communication with multi-bit rate selection is proposed in \cite{2022Adaptive} to ensure the correct information delivery at the cost of minimal number of bits. It has been observed that the feature maps usually have bell-shaped distributions with long tails \cite{2020Post,2018Value,8806964}. This means that linear quantization, which assigns the same number of quantization levels to each range, may not be able to accurately approximate the feature maps.

Another key challenge for the device-edge cooperative inference is that devices have limited computing power in general. Accordingly, the semantic encoder deployed at the device needs to be light-weight, while a more powerful semantic decoder can be deployed at the edge. The network redundancy and weight resolution have been reduced in \cite{XieHuiqiang2021A} to make the IoT devices affordable. A two-step pruning scheme is proposed in \cite{9311939} to reduce both the computation complexity and communication cost. A novel generalized divisive normalization approach \cite{2016Density} is exploited in \cite{2020Joint} to reduce the redundancy of the encoding networks. These works mainly focus on the computing power limits of devices. However, the wireless channel conditions are constantly changing, and inference tasks may have different levels of difficulty. How to design a flexible semantic communication network that can be adaptive to both the varying channel conditions and the input source is still an open question.

To address the above issues, in this paper, we propose a DL-enabled digital semantic communication system for device-edge cooperative inference, as shown in Fig. \ref{fig:system}. The goal is to maximize the accuracy of inference tasks under the constraints of communication bandwidth and computing capabilities of devices. Here, we mainly focus on the image classification application. In specific, we propose a novel digital semantic communication system with \emph{adaptive split point} for both the encoding and decoding networks. We convert features into those represented by discrete numerical values through a non-linear quantization module with trained levels, and the smaller quantization error brought by the non-linear quantization better transforms analog communication into \emph{digital communication}. Structured pruning is incorporated to reduce the number of semantic feature maps to be transmitted. Furthermore, to enable end-to-end training for the semantic communication purpose, we propose a novel \emph{semantic learning loss (SLL)} function to reduce the semantic error rather than the bit error. We also found in several networks, such as ResNet \cite{Identity} and Swin Transformer \cite{swin}, that a deeper semantic encoder does not always achieve better classification accuracy for all categories of images. As the fixed encoder design cannot flexibly deal with varying channel conditions and different images, we design a \emph{policy network} to optimize the configurations of the semantic encoding and decoding networks, by maximizing the classification accuracy under the constraint of the limited computing and bandwidth resources. Our experimental results show that the proposed semantic communication network achieves a classification accuracy close to that of deep joint source-channel coding (DJSCC), which can be seen as the upper bound of the digital semantic system \cite{9998051}. Additionally, we show the effectiveness of the proposed modules, including the non-linear quantization, structured pruning, and SLL, through ablation studies.

Our major contributions are summarized as follows:
\begin{itemize}
\item[$\bullet$] We design a light-weight and communication-efficient digital semantic communication system. For digital semantic transmission, we convert features into those represented by discrete numerical values through a non-linear quantization module with trained levels, and the smaller quantization error brought by the non-linear quantization better transforms analog communication into \emph{digital communication}. Besides, by further incorporating structured pruning in the encoding network, the communication bandwidth is reduced significantly.

\item[$\bullet$] To enable end-to-end training for the semantic communication purpose, we propose an SLL function that accurately conveys semantic information and reduces semantic error rather than bit error. The SLL function is designed to learn the ``soft'' semantic information against channel noise more effectively..

\item[$\bullet$] To enhance the flexibility of semantic transmission, we introduce a policy network that dynamically optimizes the configurations of the semantic encoding and decoding networks. The proposed policy network is designed to adaptively select the split point and maximize classification accuracy under the computing resource constraints of the device, diverse channel conditions, and images.

\item[$\bullet$] Through extensive simulations conducted on the CIFAR-10 and ImageNet datasets, it has been observed that our proposed digital semantic communication system closely approaches the performance of analog semantic communication systems. Furthermore, the results indicate that the proposed system exhibits a substantial enhancement in performance when compared to conventional method.
\end{itemize}

The rest of this paper is organized as follows. Section II presents the system model for the digital semantic communication with adaptive network split. Section III introduces the details of the system model including the transmission residual block (TRB), which consists of structured pruning module and non-linear quantization module, the SLL and the policy network. In Section IV, the proposed method is evaluated to demonstrate its validity and adaptability. Finally, Section V concludes our work.

It should be mentioned that a part of the work, mainly in Sec. $\text {III-A}$, has been presented in IEEE VTC2023-Spring \cite{10200355}. This work has significant extensions including the new network structure which exploits a policy network, and the new loss function, i.e., SLL, which exploits a semantic loss inspired by knowledge distillation.

Notations: Vectors and Matrices are denoted by boldface lower- and upper-case letters, respectively. The $i$-th element in the vector $\boldsymbol{x}$ is represented by $\boldsymbol{x}[i]$. $\mathbb{R}$ denotes the sets of real numbers. Finally, the symbol $\cdot$ denotes multiplication between real numbers and matrices, while the symbol $\otimes$ represents the multiplication of matrices by multiplying their elements at corresponding positions.

\begin{figure*}
\centering
\includegraphics[width=\linewidth]{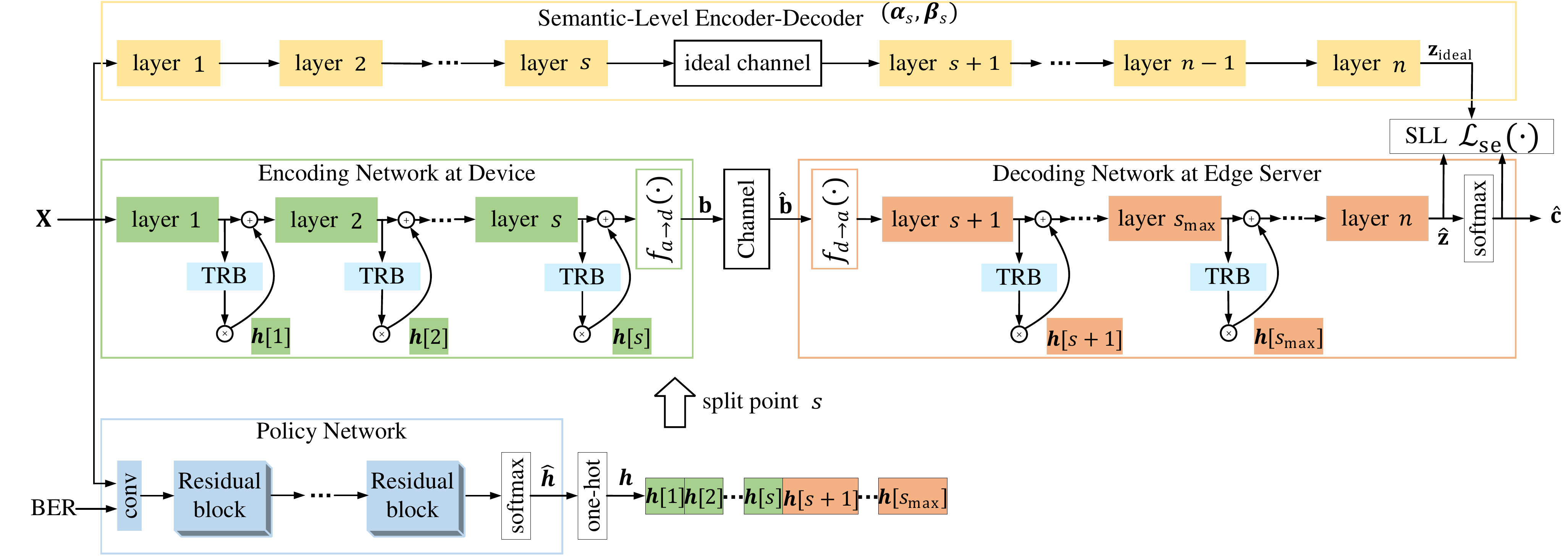}
\caption{The proposed device-edge adaptive semantic communication system. The semantic-level encoder-decoder is only considered during training. The TRB converts the semantic features into bit sequences. The policy network outputs a one-hot vector $\boldsymbol{h}$ to indicate the optimal split point $s$. For example, when $\boldsymbol{h} = [0, 0, 1, 0, \ldots, 0, 0], s=3$, the third split point is chosen by the policy network, and the third TRB after layer $3$ is activated.}
\label{fig:framework}
\end{figure*}

\section{System Model and Problem Formulation}
We consider a joint device-edge semantic communication system in the wireless network, and image classification as the target application, while the framework can be straightforwardly extended to other tasks. The device and the edge server have certain background knowledge for semantic information extraction, which is related to the application. They aim to complete inference tasks, such as image classification, in a cooperative manner. Due to limited computing power and memory, the device as the transmitter typically cannot perform all the computations required by the complex inference task. A light-weight semantic encoding network needs to be deployed to ensure that its complexity does not exceed the computing power of the device, while reducing the bandwidth requirement by efficiently extracting the semantic information at the same time.

The proposed device-edge adaptive semantic communication system is shown in Fig. \ref{fig:framework}, which has two encoder/decoder pairs to carry out the distillation of knowledge. The \emph{semantic level} encoder/decoder aims to guarantee that semantic information encoded by the resource-limited device is understood correctly by the semantic decoder at the edge. The \emph{transmission level} encoder/decoder aims to reduce the bandwidth cost for transmitting the semantic information. The encoding network deployed at the transmitter extracts the semantic features from the input image $\mathbf{X}$, and then converts them to bit sequences. The received semantic information needs to be understood at the receiver by the semantic decoding network. To achieve successful recovery at the semantic level, the ``soft'' semantic information $\mathbf{z}_{\text{ideal}}$ that contains the ``soft'' semantic information, is used to minimize the received semantic error. The proposed TRB exploits structured pruning and non-linear quantization techniques to reduce the bandwidth cost at the transmission level. Furthermore, a \emph{policy network} is employed to optimize the configurations of the encoding and decoding networks in this system. The goal is to maximize the classification accuracy under the bandwidth limit and the computing capability constraint of the device.

The \emph{semantic level} DNN for inference in the ideal channel environment is denoted by $\mathbf{\Psi}$, with totally $n$ layers. This DNN $\mathbf{\Psi}$ is split into two sub-networks $\boldsymbol{\alpha}_{s}$ and $\boldsymbol{\beta}_{s}$, which act as the semantic-level encoder and decoder, respectively. Here, $s$ is the split point optimized by the policy network, meaning that $\boldsymbol{\alpha}_{s}$ has the former $s$ layers of $\mathbf{\Psi}$. The encoding network is represented by a trainable network $S^{(e)}(\cdot;\boldsymbol{\alpha}_{s},\boldsymbol{\delta}_{s},\boldsymbol{\theta}_{s})$, which consists of a semantic encoder represented as $E(\cdot;\boldsymbol{\alpha}_{s})$, and the proposed TRB. Within the TRB, there is a structured pruning module $A(\cdot;\boldsymbol{\delta}_{s})$, which reduces the bandwidth cost by extracting the most important semantic features, and a quantization module $Q(\cdot;\boldsymbol{\theta}_{s})$, which maps the features in a continuous range to the trained quantization levels for digital transmission. Here, $\boldsymbol{\alpha}_{s}$, $\boldsymbol{\delta}_{s}$ and $\boldsymbol{\theta}_{s}$ are the trainable parameters of the encoder, the structured pruning and the quantization module, respectively. The whole encoding process on the device can be written as:
\begin{equation}
\mathbf{S}_{e} = S^{(e)}(\mathbf{X};\boldsymbol{\alpha}_{s},\boldsymbol{\delta}_{s},\boldsymbol{\theta}_{s}) =Q\left(A\left(E\left(\mathbf{X};\boldsymbol{\alpha}_{s}\right);\boldsymbol{\delta}_{s}\right);\boldsymbol{\theta}_{s}\right),
\end{equation}
where $\mathbf{X}\in \mathbb{R}^{w_{1} \times h_{1} \times 3}$ and $\mathbf{S}_{e} \in \mathbb{R}^{w_{2} \times h_{2} \times z}$ are the input image data and the semantic feature, respectively. $w_{2} \times h_{2}$ is the dimension of a feature map, and $z$ is the number of channels of the semantic features. The output feature of the semantic encoder is quantized to $2^{q}$ discrete quantization levels, and mapped to a bit sequence $\mathbf{b} \in \{0,1\}^{w_{2}h_{2}zq}$. Each trainable quantization level represents $q$ bits in the sequences. That is represented as $\mathbf{b}=f_{a \to d}(\mathbf{S}_{e},\boldsymbol{\theta}_{s})$. 

We consider a binary symmetric channel (BSC). At the edge server, the received bit sequences are denoted by $\hat{\mathbf{b}}\in \{0,1\}^{w_{2}h_{2}zq}$, and mapped to the trained $2^{q}$ quantization levels. This is represented as $\hat{\mathbf{S}}_{e}=f_{d \to a}(\hat{\mathbf{b}}; \boldsymbol{\theta}_{s})$. The decoder network understands the semantic information with transmission impairments via training under the novel SLL function. The semantic decoder can be represented by a trainable decoder network $S^{(d)}(\cdot;\boldsymbol{\beta}_{s})$, where $\boldsymbol{\beta}_{s}$ is the trainable parameters of the decoder. At the receiver side, the semantic features are reconstructed, and the remaining part of the forward pass is completed to obtain the final prediction, which can be denoted by
\begin{equation}
\hat{\mathbf{c}}=S^{(d)}\left(\hat{\mathbf{S}}_{e};\boldsymbol{\beta}_{s}\right),
\end{equation}
where $\hat{\mathbf{c}}$ represents the output of the DNN at the receiver, i.e., the classification results.

\begin{figure}
\centering
\includegraphics[width=0.5\linewidth]{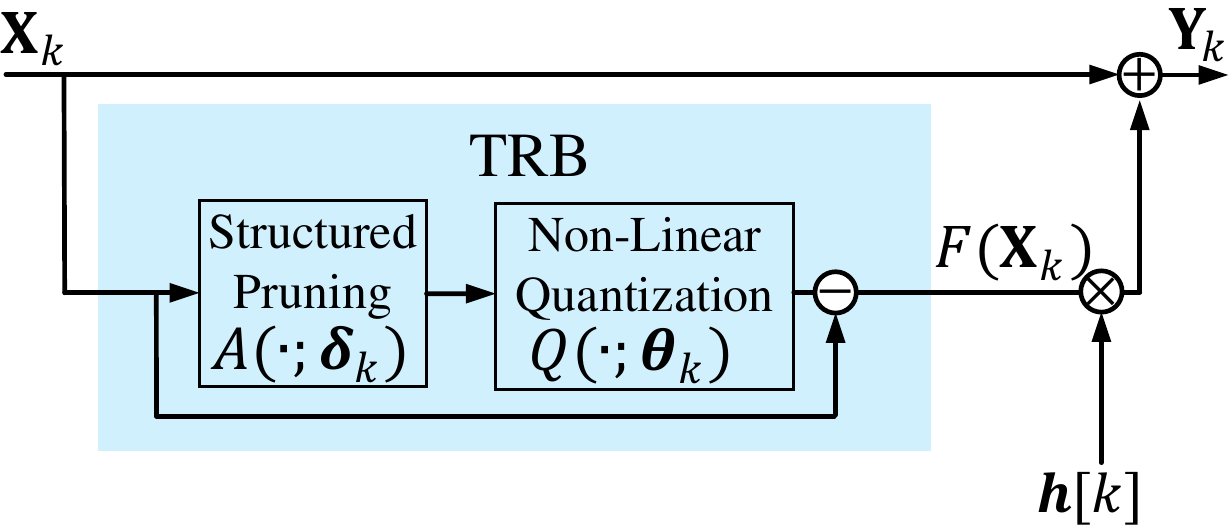}
\caption{The architecture of the TRB.}
\label{fig:TRB}
\end{figure}

The policy network $P(\cdot;\boldsymbol{\varphi})$ optimizes the configuration of the encoding network and decoding network, and aims to maximize the classification accuracy. To achieve this goal, the policy network $P(\cdot;\boldsymbol{\varphi})$ uses the image data and the bit error rate (BER) to adjust the split point, resulting in a split point that achieves maximum classification accuracy, which can be represented as
\begin{equation}
\hat{\boldsymbol{h}} = P(\mathbf{X},\text{BER};\boldsymbol{\varphi}),
\end{equation}
where $\boldsymbol{\varphi}$ is the trainable parameters of the policy network, which gives the probability distribution of split point by the softmax. The point with the maximum probability is selected as the split point. As shown in Fig. \ref{fig:framework}, $\boldsymbol{h} \in \{0,1\}^{s_{\text{max}}}$ is a one-hot value, which is obtained based on the result of the softmax operation, $\hat{\boldsymbol{h}}$, where $\boldsymbol{h}[s]=1$ if $s$ is the index of the split point with the maximum probability, i.e., $\hat{\boldsymbol{h}}[s]=\max _{k \in \{1,2,\ldots,s_{\text{max}}\}}(\hat{\boldsymbol{h}}[k])$. $s_{\text{max}}$ represents the largest number of layers supported by the resource-limited device, which can be expressed as:
\begin{equation}
f({\boldsymbol{\alpha}_{s_{\text{max}}},\boldsymbol{\delta}_{s_{\text{max}}},\boldsymbol{\theta}_{s_{\text{max}}}}) \leqslant f_{\text{max}}^{\text{dev}} < f({\boldsymbol{\alpha}_{s_{\text{max}}+1},\boldsymbol{\delta}_{s_{\text{max}}+1},\boldsymbol{\theta}_{s_{\text{max}}+1}}),
\end{equation}
where $f(\cdot)$ and $f_{\text{max}}^{\text{dev}}$ represent the required floating-point operations (FLOPs) of the encoding network and the maximum FLOPs supported by the device, respectively. The one-hot vector $\boldsymbol{h}$ indicates the optimal split point $s$. The corresponding TRB is activated, and the semantic features are 
quantized to the $2^{q}$ trained quantization levels, where each element of the features is converted to $q$ transmitted bits according to $\boldsymbol{h}$. As shown in Fig. \ref{fig:TRB}, the TRB can be represented as:
\begin{equation}
\begin{aligned}
\mathbf{Y}_{k} &=\mathbf{X}_{k}+\boldsymbol{h}[k] F\left(\mathbf{X}_{k}\right), \\
F\left(\mathbf{X}_{k}\right) &=Q(A\left(\mathbf{X}_{k};\boldsymbol{\delta}_{k}\right);\boldsymbol{\theta}_{k})-\mathbf{X}_{k}, \forall k \leqslant s_{\text{max}},
\end{aligned}
\end{equation}
where $\mathbf{Y}_{k}, \mathbf{X}_{k}, F\left(\mathbf{X}_{k}\right) \in \mathbb{R}^{w_{2} \times h_{2} \times z}$. $\mathbf{Y}_{k}$ is the potential transmission features at the $k$-th split point. $\mathbf{X}_{k}$ and $F\left(\mathbf{X}_{k}\right)$ are input and output of the $k$-th TRB, respectively. $\boldsymbol{h}[k]$ represents the activation of the $k$-th TRB, which is decided by the policy network. $Q\left(\cdot;\boldsymbol{\theta}_{k}\right)$ and $A\left(\cdot;\boldsymbol{\delta}_{k}\right)$ are the quantization module and the structured pruning in the $k$-th TRB, respectively. For example, when $\boldsymbol{h} = [0, 0, 1, 0, \ldots, 0, 0], s=3$, the third split point is chosen by the policy network. The third TRB after layer $3$ is activated. $\mathbf{S}_{e} = \mathbf{Y}_{3} = Q(A\left(\mathbf{X}_{3};\boldsymbol{\delta}_{3}\right);\boldsymbol{\theta}_{3})$, while $\mathbf{Y}_{k}=\mathbf{X}_{k}$ for $\forall k \neq 3$. The semantic encoder $E(\mathbf{X};\boldsymbol{\alpha}_{3})$ on the device consists of layer $1$ to layer $3$. The details of the policy network will be introduced in Sec. III-D.
 
More complex encoding networks usually have stronger semantic extraction capabilities, but devices with limited computing resources are difficult to afford. In addition, the entire semantic system is expected to achieve the best performance under the limited bandwidth. The optimization loss function can be represented as:
\begin{equation}
\begin{aligned}
\min _{\boldsymbol{\alpha}_{s},\boldsymbol{\beta}_{s},\boldsymbol{\delta}_{s},\boldsymbol{\theta}_{s},s}& \mathcal{L}_{\text{se}}(\boldsymbol{\alpha}_{s},\boldsymbol{\beta}_{s},\boldsymbol{\delta}_{s},\boldsymbol{\theta}_{s}) + \gamma \mathcal{L}_{\text{c}}(\boldsymbol{\delta}_{s})\\
 \text { s.t. }\quad&  1 \leqslant s \leqslant s_{\text{max}}, s\in \mathbb{Z}, \\
& w_{2}h_{2}zq \leqslant B_{\text{max}}, \\
\label{problem}
\end{aligned}
\end{equation}
where $\mathcal{L}_{\text{se}}(\boldsymbol{\alpha}_{s},\boldsymbol{\beta}_{s},\boldsymbol{\delta}_{s},\boldsymbol{\theta}_{s})$ is the proposed SLL function, which will be introduced in Sec. $\text {III-C}$. $\mathcal{L}_{\text{c}}(\boldsymbol{\delta}_{s})$ is used to minimize the bandwidth required for semantic information transmission. The parameter $\gamma$ balances the trade off between the system performance and bandwidth requirement. The constraints indicate that the computing power required by the semantic encoding network is no more than that of the device, and the number of bits cannot exceed the maximum bit $B_{\text{max}}$ per unit time.

\section{Joint Device-Edge Semantic Communication System}

In this section, we introduce in detail the joint device-edge digital semantic communication system, which consists of encoding network, decoding network and policy network. The encoding network, which has the semantic encoder, structured pruning and module non-linear quantization module, extracts semantic features and converts them into the features with several discrete quantization levels. The decoding network reconstructs and understands the semantic information with transmission impairments by the novel SLL function. The policy network aims to optimize the configuration of the encoder and decoder given the resource limits, channel conditions, e.g., BERs, and the images.

\subsection{The TRB in the Encoding Network}
The TRB has a structured pruning module for reducing the bandwidth requirement and a non-linear quantization module, which converts continuous semantic features into discrete features with several trained quantization levels.

\subsubsection{The Structured Pruning Module}
\begin{figure}
\centering
\includegraphics[width=0.5\linewidth]{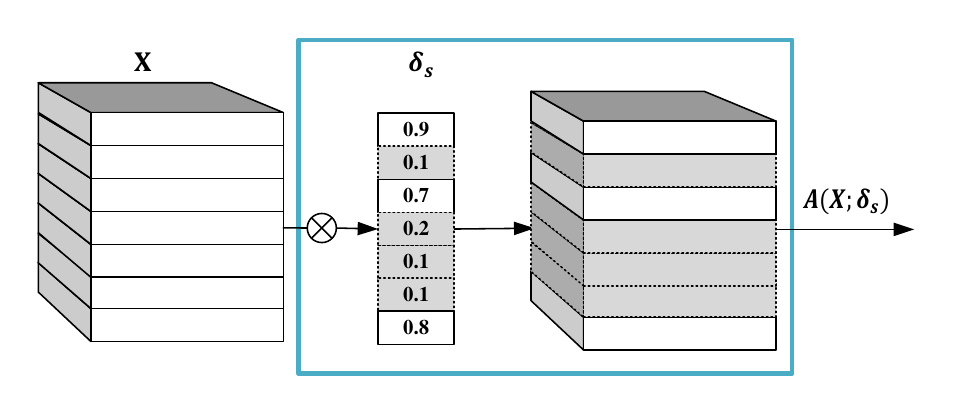}
\caption{An illustration of the structured pruning in the semantic communication system. The feature map in gray is pruned, and the white one is retained.}
\label{fig:atten}
\end{figure}


One important goal on designing a joint device-edge communication system is to reduce the number of bits to be transmitted. Nevertheless, as the number of channels is usually expanded to extract the most significant features from the input image, the number of the intermediate semantic features is higher than that of the input in most popular convolutional neural networks (CNNs), such as ResNet \cite{Deep}, and Transformer \cite{2017Attention}. Therefore, we incorporate a structured pruning to reduce the redundancy of the semantic features. As shown in Fig. \ref{fig:atten}, this module learns to keep only the most valuable semantic features, thus reducing the bandwidth requirement. 

In specific, we introduce a scaling vector $\boldsymbol{\delta}_{s} \in \mathbb{R}^{z}$. We incorporate the $l_{1}$-regularization function to enforce the scaling parameter to be sparse during training. Therefore, the loss function of this goal can be formulated as:
\begin{equation}
\mathcal{L}_{\text{c}}(\boldsymbol{\delta}_{s}) = \left\|\boldsymbol{\delta}_{s}\right\|_{1},
\end{equation}
where $\left\|\boldsymbol{\cdot}\right\|_{1}$ is the $l_{1}$-norm and widely used to enforce sparsity. When the elements in the scaling vector are smaller than the given threshold, the corresponding channels of the semantic features are clipped out, i.e., they are not transmitted to the edge server. As shown in Fig. \ref{fig:atten}, for $\forall k \in \{1,\ldots, z\}$, if the scaling vector is smaller than a given threshold $\eta$, i.e., $\boldsymbol{\delta}_{s}[k]<\eta$, then the corresponding feature map is clipped out. This process is denoted as $A(\cdot;\boldsymbol{\delta}_{s})$. We leverage the scaling in the structured pruning to effectively identify the most valuable semantic features on the device.
\subsubsection{The Non-Linear Quantization Module}

\begin{figure*}[htp!]
    \centering
    \subfigure[Ideal quantization function]{\label{4c}\includegraphics[width=0.32\hsize, height=0.27\hsize]{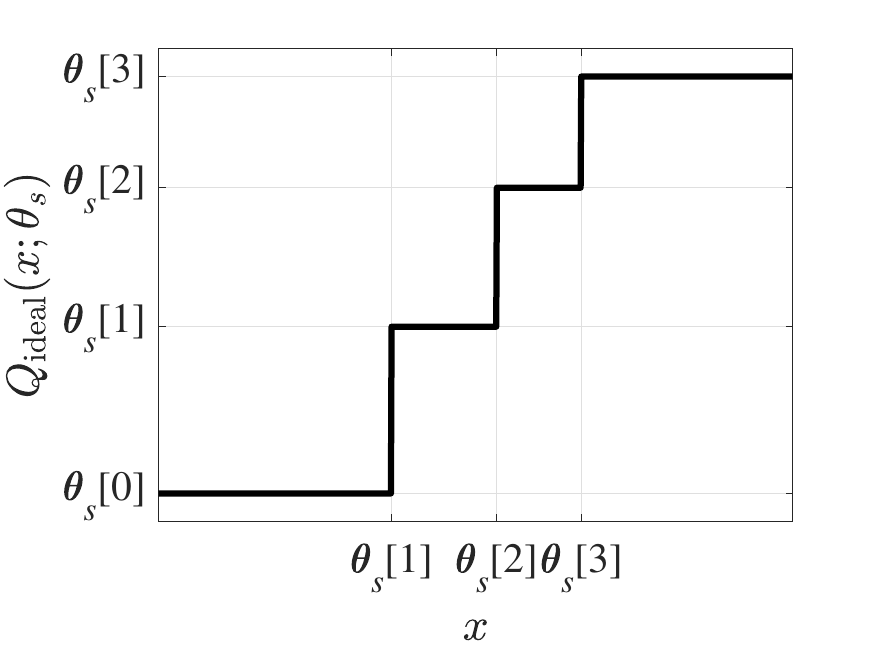}}
\label{nonquantization}
    \subfigure[$T$=10]{\label{4d}\includegraphics[width=0.32\hsize, height=0.27\hsize]{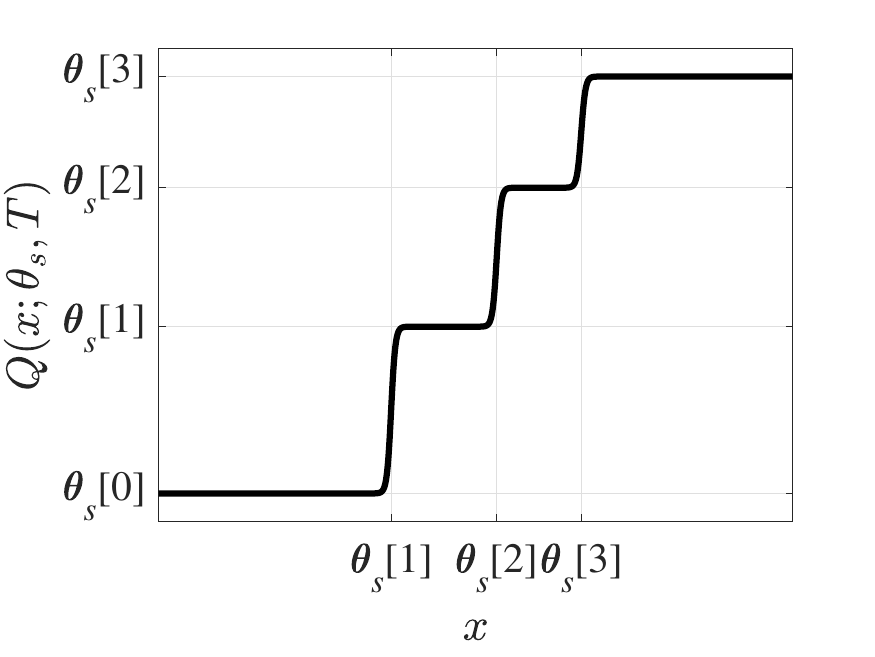}}
    \subfigure[$T$=100]{\label{4f} \includegraphics[width=0.32\hsize, height=0.27\hsize]{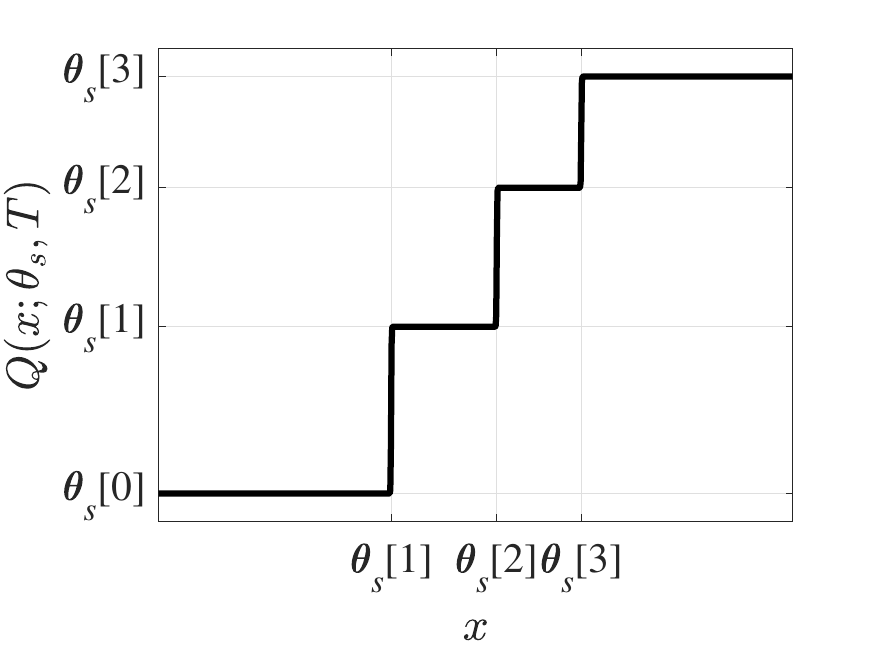}}
    \caption{The ideal non-linear quantization function and the proposed quantization function. During the training stage, the ``soft'' quantization function goes from a straight line to steps as the $T$ increases.}
    \label{quantization}
\end{figure*}

Inspired by \cite{2019Quantization}, it converts floating-point values into integers that can be more effectively represented in binary format to facilitate more efficient binary operations. However, quantizing to fixed integers holds no significance in terms of communication and results in additional computational operations. Additionally, quantizing these values to fixed integers is not beneficial for the neural network's ability to learn better features. Therefore, we formulate the quantization operation that is more suitable for communication. Our proposed quantization function translates floating-point values into the maximum value among adjacent fixed two values learned by the network. This function is represented as a linear combination of several sigmoid functions with trained values, i.e., the quantization levels. An approximate expression is learned via continuous relaxation of the sigmoid functions, allowing for differentiability.

The quantization operation maps continuous inputs into discrete quantization levels. Given input $\mathbf{X}$, the conventional linear quantization scheme can be expressed as:
\begin{equation}
Q(\mathbf{X})=\frac{1}{q_{w}} \cdot \operatorname{round}\left(q_{w} \cdot \left(\mathbf{X}-\min \left(\mathbf{X}\right)\right)\right) + \min \left(\mathbf{X}\right),
\end{equation}
where $q_{w}=\frac{2^{q}-1}{\max \left(\mathbf{X}\right)-\min \left(\mathbf{X}\right)}$ is a scaling factor to map the dynamic range of float points to the discrete numbers, and the output can be represented as a $q$-bit symbol. $\operatorname{round}(\cdot)$ is the rounding operation. However, many studies have shown that the distributions of weights and activation of DNNs are similar to bell-shaped Gaussian or Laplacian \cite{2020Post,2016Deep}. Therefore, the linear quantization typically results in more quantization error than the non-linear quantization, leading to the features distortion, especially in low-bit cases. In addition, the low-bit quantization of networks is formulated as an approximation problem in these methods, which confronts the gradient mismatch problem since the used quantization function is non-differentiable \cite{2019Quantization}.

To deal with the drawback of linear quantization, we propose a novel non-linear quantization that has \emph{dynamic quantization levels learned by the network itself}. Firstly, the ideal non-linear quantization function can be represented as a combination of several unit step functions with specified biases, which can be represented as:
\begin{equation}
\begin{aligned}
&Q_{\text{ideal}}(\mathbf{X}_{s};\boldsymbol{\theta}_{s}) \\
=\ &\boldsymbol{\theta}_{s}[0] \cdot \boldsymbol{1} + \sum_{i=1}^{2^{q}-1}(\boldsymbol{\theta}_{s}[i]-\boldsymbol{\theta}_{s}[i-1]) \cdot \mu\left(\mathbf{X}_{s}-\boldsymbol{\theta}_{s}[i] \cdot  \boldsymbol{1} \right),
\label{Q_ideal}
\end{aligned}
\end{equation}
where $\boldsymbol{1} \in \mathbb{R}^{w_{2} \times h_{2} \times z}$ is an array where each element equals one. $\mu(\cdot)$ is the unit step function. $\boldsymbol{\theta}_{s}\in \mathbb{R}^{2^{q}}$, $(\boldsymbol{\theta}_{s}[i]-\boldsymbol{\theta}_{s}[i-1])$ and $\boldsymbol{\theta}_{s}[i]$ are the scales and the bias of the $i$-th unit step function, respectively, and $\boldsymbol{\theta}_{s}[0]$ is the first quantization level. Note that the subscript $s$ is denoted as the quantization module in the $s$-th TRB. Fig. \ref{4c} shows the ideal non-linear quantization function. Since the standard unit step function is not differentiable, previous methods \cite{2018JALAD,2019BottleNet,XieHuiqiang2021A,2016DoReFa} use a straight-through estimator to estimate the gradient of quantized features in the back-propagation. However, using different forward and backward approximations results in the gradient mismatch problem, which makes the optimization unstable \cite{2019Quantization}. 

Since the unit step function is not smooth, we then adopt a continuous relaxation method for training, by replacing each non-differentiable unit step function in the ideal quantization function with a differentiable sigmoid function \cite{2019Quantization}. However, the difference between the step function and the sigmoid function cannot be ignored. To solve this issue, a factor $T$ is introduced to weaken this difference. The continuous relaxation of the sigmoid function with $T$ is proposed to replace the unit step function, which can be derived as:
\begin{equation}
\sigma(x;T)=\frac{1}{1+\exp (-Tx)},
\label{Q_base}
\end{equation}
where the gap between the unit step function and the relaxed sigmoid function becomes smaller as $T$ gets larger, as shown in Fig. \ref{4d}-\ref{4f}. Such a relaxed sigmoid function is differentiable, and the gradient of the relaxed sigmoid function in the back-propagation can be represented as:
\begin{equation}
\frac{\partial \sigma(x;T)}{\partial x} = T \sigma(x;T)(1-\sigma(x;T)).
\end{equation}

The ideal quantization function can be formulated as the sum of several unit step functions as shown in Eq. (\ref{Q_ideal}). By replacing each unit step function with Eq. (\ref{Q_base}), the proposed ``soft'' quantization function can be represented as:
\begin{equation}
Q\left(\mathbf{X}_{s};\boldsymbol{\theta}_{s},T\right)=\boldsymbol{\theta}_{s}[0] \cdot  \boldsymbol{1} + \sum_{i=1}^{2^{q}-1} (\boldsymbol{\theta}_{s}[i]-\boldsymbol{\theta}_{s}[i-1]) \cdot \sigma_{i}\left(\mathbf{X}_{s};T\right),
\label{Q}
\end{equation}
where $\sigma_{i}(\mathbf{X}_{s};T)=\sigma\left(\mathbf{X}_{s}-\boldsymbol{\theta}_{s}[i] \cdot  \boldsymbol{1};T\right)$. In particular, $\boldsymbol{\theta}_{s}=[\boldsymbol{\theta}_{s}[0],\boldsymbol{\theta}_{s}[1],...,\boldsymbol{\theta}_{s}[2^{q}-1]]$ includes the parameters to be learned. The feature $\mathbf{S}_{e}$ is mapped to bit sequences, i.e., $\mathbf{b}=f_{a \to d}(\mathbf{S}_{e};\boldsymbol{\theta}_{s})$. Each quantization level represents $q$ bits in the sequences. For example, when $q=2$, $f_{a \to d}(\cdot ; \boldsymbol{\theta}_{s})$ is defined by:
\begin{equation}
\begin{aligned}
f_{a \to d}(x ;\boldsymbol{\theta}_{s})=
&\left\{\begin{aligned}
11 &~~~ x > \frac{\boldsymbol{\theta}_{s}[2] + \boldsymbol{\theta}_{s}[3]}{2} \\
10 &~~~ \frac{\boldsymbol{\theta}_{s}[1] + \boldsymbol{\theta}_{s}[2]}{2} < x \leq \frac{\boldsymbol{\theta}_{s}[2] + \boldsymbol{\theta}_{s}[3]}{2} \\
01 &~~~ \frac{\boldsymbol{\theta}_{s}[0] + \boldsymbol{\theta}_{s}[1]}{2} < x \leq \frac{\boldsymbol{\theta}_{s}[1] + \boldsymbol{\theta}_{s}[2]}{2} \\ 
00 &~~~ x \leq \frac{\boldsymbol{\theta}_{s}[0] + \boldsymbol{\theta}_{s}[1]}{2}
\end{aligned}\right.
\end{aligned}
\end{equation}

We considered that using a larger $T$ to generate discrete quantization features would degrade the expressive power of the network at the beginning of the training stage, so we start with a small $T$ to ensure stable and effective learning, and gradually increase $T$ w.r.t. the training epochs to finally approach the ideal quantization functions. Besides, we need to back-propagate the gradients of the loss through the proposed quantization function, as well as computing the gradients with respect to the involved parameters:

\begin{equation}
\begin{aligned}
&\frac{\partial  Q\left(\mathbf{X}_{s};\boldsymbol{\theta}_{s},T\right)}{\partial \mathbf{X}_{s}} = \\
& T \sum_{i=1}^{2^{q}-1} (\boldsymbol{\theta}_{s}[i]-\boldsymbol{\theta}_{s}[i-1]) \cdot \sigma_{i}\left(\mathbf{X}_{s};T\right) \otimes \left(\boldsymbol{1}-\sigma_{i}(\boldsymbol{x};T)\right),\\
\end{aligned}
\end{equation}

\begin{equation}
\begin{aligned}
&\frac{\partial Q\left(\mathbf{X}_{s};\boldsymbol{\theta}_{s},T\right)}{\partial \boldsymbol{\theta}_{s}[i]}=  \\
&\left\{\begin{aligned}
\boldsymbol{1}-&\sigma_{1}(\mathbf{X}_{s};T),~~~~~~~~~~~~~~~~~~~~~~~~~~i=0 \\
-T&\left(\boldsymbol{\theta}_{s}[i]-\boldsymbol{\theta}_{s}[i-1]\right) \cdot \sigma_{i}(\mathbf{X}_{s};T) \otimes \left(\boldsymbol{1}-\sigma_{i}(\mathbf{X}_{s};T)\right) \\
&+\sigma_{i}(\mathbf{X}_{s};T)-\sigma_{i+1}(\mathbf{X}_{s};T),~~~~0<i<2^{q}-1 \\ 
-T&\left(\boldsymbol{\theta}_{s}[2^{q}]-\boldsymbol{\theta}_{s}[2^{q}-1]\right)\!\cdot\! \sigma_{2^{q}}(\mathbf{X}_{s};T)\!\otimes\! \left(\boldsymbol{1}-\sigma_{2^{q}}(\mathbf{X}_{s};T)\right) \\
&+\sigma_{2^{q}}(\mathbf{X}_{s};T),~~~~~~~~~~~~~~~~~~~~~i=2^{q}-1.
\end{aligned}\right.
\end{aligned}
\end{equation}

In summary, in this part, we propose a ``soft'' differentiable quantization function to approximate the ideal quantization function with a gradually increased $T$ according to Eq. (\ref{Q}). The proposed non-linear quantization function has no approximation of gradients, and thus avoids gradient mismatch during training. Moreover, it can effectively train the quantization levels to better adapt to the features with bell-shaped distribution.

\subsection{Feature Reconstruction through BSC}
The feature $\mathbf{S}_{e}$ is quantizated by the $f_{a \to d}(\cdot;\boldsymbol{\theta}_{s})$, and mapped to the transmitted sequence $\mathbf{b}$, which goes through a BSC channel and is received by the receiver as $\hat{\mathbf{b}}$ with error bits. At the edge server, $\hat{\mathbf{b}}$ is mapped to the trained $2^{q}$ quantization levels, and the received feature $\hat{\mathbf{S}}_{e}$ is reconstructed via $f_{d \to a}(\cdot ; \boldsymbol{\theta}_{s})$. When $q=2$, $f_{d \to a}(\cdot;\boldsymbol{\theta}_{s})$ is defined by:
\begin{equation}
f_{d \to a}(x;\boldsymbol{\theta}_{s})= \begin{cases}
\boldsymbol{\theta}_{s}[3] & x = 11 \\ 
\boldsymbol{\theta}_{s}[2] & x = 10 \\ 
\boldsymbol{\theta}_{s}[1] & x = 01 \\ 
\boldsymbol{\theta}_{s}[0] & x = 00.\end{cases}
\end{equation} 

\subsection{End-to-End Training with SLL}

In this subsection, we propose a novel SLL function to minimize the semantic error. Recall that in semantic communications, semantic information is transmitted after extracting the meanings of data and filtering out the irrelevant or unessential information. The transmitted semantic meaning is extracted by the encoding network, and the received meaning is then understood by the decoding network. Different from BER, the semantic error loss can be regarded as the gap between the real meaning and that understood by the receiver. Inspired by knowledge distillation \cite{2015Distilling,2018Improving}, the ``soft'' information $\mathbf{z}_{\text{ideal}}$, i.e., the output of last layer in a DNN, with the ideal channel is known as ``soft'' semantic information. The semantic gap between $\mathbf{z}_{\text{ideal}}$ and $\hat{\mathbf{z}}$ is used to minimize the semantic error. Thus, the gap of the ``soft'' information between the network with ideal channel and the joint edge-device network is considered in the SLL. The proposed SLL can be represented as follows:
\begin{table}[t]
\begin{center}
\caption{An intuitive example of the temperature factor $T_{\text{se}}$ and the distribution of the semantic information. $10^{x}$ is represented as $e^{x}$.}
\begin{tabular}{c|c|c|c|c|c}
\hline
 Predictions & Cow & Dog & Cat & Car & Person \\
\hline
 ${z}_{\text{ideal}}$ & $3.39$ & $18.26$ & $13.59$ & $-5.12$ & $-7.35$ \\
\hline
 \!\!\!\!Soft targets\! ($T_{\text{se}}\!\!\rightarrow\!\!0$)\!\!\!\!& $0.2$ & $0.2$ & $0.2$ & $0.2$ & $0.2$ \\
 \!\! Soft targets \!($T_{\text{se}}\!\!=\!\!1$)\!\! & $0.11$ & $0.49$ & $0.31$ & $0.05$ & $0.04$ \\  
 \!\!\!\!Soft targets \!($T_{\text{se}}\!\!=\!\!10$)\!\!\!\!&\!\!$3.4e^{-7}\!\!$ &\!\!$0.99$\!\!&\!\!$9.2e^{-3}$\!\!&\!\!$6.9e^{-11}$\!\!&\!\!$7.6e^{-12}$\!\! \\  
 \!\!\!\!Hard targets \!($T_{\text{se}}\!\!\rightarrow\!\!\infty$)\!\!\! & $0$ & $1$ & $0$ & $0$ & $0$\\  
\hline
\end{tabular}
\label{tab:example}
\end{center}
\end{table}

\begin{equation}
\begin{aligned}
&\mathcal{L}_{\text{se}}(\boldsymbol{\alpha}_{s},\boldsymbol{\beta}_{s},\boldsymbol{\delta}_{s},\boldsymbol{\theta}_{s})\\ 
=\ &\lambda  F_{\mathrm{CE}}\left(\mathbf{c}, \hat{\mathbf{c}}\right)+(1-\lambda) F_{\text{se}}\left(\mathbf{z}_{\text{ideal}}, \hat{\mathbf{z}}; T_{\text{se}}\right),
\end{aligned}
\end{equation}
where $F_{\mathrm{CE}}\left(\mathbf{c}, \hat{\mathbf{c}}\right) = - \sum_{j \in \Omega}^{l} c[j] \log \hat{c}[j] + (1-c[j]) \log (1-\hat{c}[j])$ is the cross-entropy (CE) loss function. $\mathbf{c}\in \{0,1\}^{l}$ and $\hat{\mathbf{c}} \in \mathbb{R}^{l}$ are the ground truth and the predicted output, respectively. $\mathbf{z}_{\text{ideal}}, \hat{\mathbf{z}} \in \mathbb{R}^{l}$. $l$ is the number of classes. $\Omega$ denotes the set of classes. $c[j]$ and $\hat{c}[j]$ are the $j$-th elements of $\mathbf{c}$ and $\hat{\mathbf{c}}$, respectively. $c[j] \in \{0,1\}$ is the real probability of the $j$-th class, and $\hat{c}[j]$ is the predicted probability of the $j$-th class. $F_{\text{se}}$ is a knowledge distillation loss, which is used to learn semantic information to reduce the semantic gap. The ``soft'' semantic information $\hat{\mathbf{z}}$ can be used to calculate the predicted class $\hat{\mathbf{c}}$ by a softmax function:
\begin{equation}
\hat{\mathbf{c}}=softmax(\hat{\mathbf{z}}),
\end{equation}
where $\hat{c}[j]=\frac{\exp \left(\hat{z}[j]\right)}{\sum_{j \in \Omega} \exp \left(\hat{z}[j]\right)}$. $\hat{z}[j]$ is the $j$-th element of $\hat{\mathbf{z}}$. The $\mathbf{z}_{\text{ideal}}$ obtained by the ideal network contains the ``soft'' semantic information and can be used as a supervisor to transfer semantic information from the transmitter to the receiver. A temperature factor $T_{\text{se}}$ is introduced to control the distribution of the semantic information, which can be expressed as:
\begin{equation}
\mathbf{I}(\mathbf{z}_{\text{ideal}};T_{\text{se}})=softmax\left(\frac{\mathbf{z}_{\text{ideal}}}{T_{\text{se}}}\right),
\end{equation}
where $\mathbf{I}(\mathbf{z}_{\text{ideal}};T_{\text{se}}) \in \mathbb{R}^{l}$. A higher temperature $T_{\text{se}}$ produces a softer distribution. Specifically, when $T_{\text{se}} \rightarrow \infty$, all classes share the same probability. When $T_{\text{se}} \rightarrow 0$, the predictions of soft targets become one-hot labels, i.e., the ground truth $\mathbf{c}$. An intuitive example about the temperature and the distribution is shown in Table \ref{tab:example}. The ``soft'' semantic information with a suitable distribution given by $T_{\text{se}}$ helps the decoding network to understand the semantic information against semantic error caused by transmission. $F_{\text{se}}(\cdot;T_{\text{se}})$ with a suitable $T_{\text{se}}$ can be represented as:
\begin{equation}
\begin{aligned}
&F_{\text{se}}\left(\mathbf{z}_{\text{ideal}}, \hat{\mathbf{z}}; T_{\text{se}}\right)  \\
=&- T_{\text{se}}^{2} \sum_{j \in \Omega} I_{j}(\mathbf{z}_{\text{ideal}};T_{\text{se}}) \log \left(\frac{I_{j}(\mathbf{z}_{\text{ideal}};T_{\text{se}})}{I_{j}(\hat{\mathbf{z}};T_{\text{se}})}\right),
\end{aligned}
\end{equation}
where $I_{j}(\mathbf{z}_{\text{ideal}};T_{\text{se}})$ and $I_{j}(\hat{\mathbf{z}};T_{\text{se}})$ are the semantic information of the ideal network of the $j$-th class and the predicted semantic information of the $j$-th class, respectively.

\subsection{The Policy Network}
\begin{figure}
\centering
\includegraphics[width=0.5\linewidth]{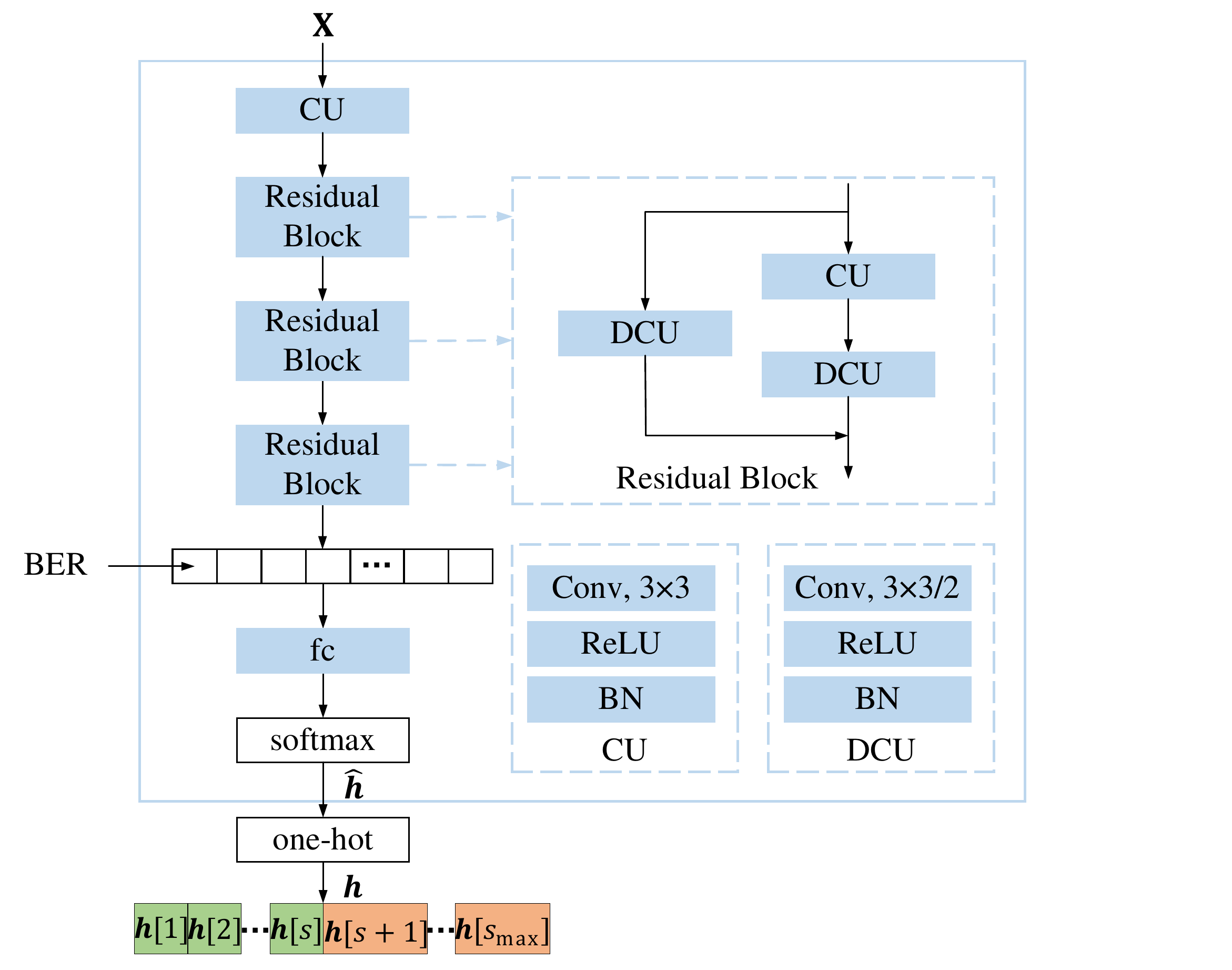}
\caption{The architecture of the policy network.}
\label{fig:policy}
\end{figure}

In this subsection, we design a policy network to adaptively optimize the configurations of the semantic encoding and decoding networks to adapt to diverse channel conditions and images. We found in several networks, such as ResNet \cite{Identity} and Swin Transformer \cite{swin} that the deeper semantic encoder can achieve better average classification accuracy, while not all categories of images can achieve better classification accuracy in the deeper encoder. The fixed encoder cannot flexibly deal with varying channel conditions and images. The split point $s$ is expected to adaptively adjust to alter the semantic encoding and decoding network, achieving the best accuracy under the constraint of the limited computing and bandwidth resources, which can be represented as:
\begin{equation}
\begin{aligned}
\max _{\boldsymbol{\alpha}_{s},\boldsymbol{\beta}_{s},\boldsymbol{\delta}_{s},\boldsymbol{\theta}_{s},s} & ACC(\boldsymbol{\alpha}_{s},\boldsymbol{\beta}_{s},\boldsymbol{\delta}_{s},\boldsymbol{\theta}_{s},s) \\
 ~~~\text { s.t. } \quad& s\in \mathbb{Z}, 1 \leqslant s \leqslant s_{\text{max}}, w_{2}h_{2}zq \leqslant B_{\text{max}}. \\
\label{acc_problem}
\end{aligned}
\end{equation}

To train the entire system, we propose a CNN named the policy network to determine the split point $s$. The architecture of the policy network is shown in Fig. \ref{fig:policy}. We use three policy residual blocks to roughly estimate the semantic information of the images, which is then concatenated with the BER. The convolution unit (CU) has a convolutional filter followed by ReLU and batch normalization (BN). The down sample convolution unit (DCU) is used to reduce the height, width and channel of the input tensor dimensions by strides and filters. The residual block consists of 3 CU and 1 residual connection. In the residual block, the output dimension of the first CU is the same as the input dimension, and the second CU and the CU in the residual connection are the DCUs. The first layer of the policy network is a CU. The following three residual blocks are used to extract the feature of configurations from the data. This feature concatenated with the BER is activated by a full connection layer following a softmax function, producing a ``soft'' configuration $\hat{\boldsymbol{h}} \in \mathbb{R}^{s_{\text{max}}}$, i.e., the probability distribution of the split point, to make an optimal choice in these limited split point. The point with the maximum probability is the split point, and producing an one-hot value $\boldsymbol{h}$ as the configuration of the encoding network. Note that the complexity of the policy network is not high, i.e., close to the complexity of one layer in Fig. \ref{fig:framework}. The goal of the proposed system with the policy network can be represented as:
\begin{equation}
\max _{\boldsymbol{\alpha}_{s},\boldsymbol{\beta}_{s},\boldsymbol{\delta}_{s},\boldsymbol{\theta}_{s},\boldsymbol{\varphi}} ACC(\boldsymbol{\alpha}_{s},\boldsymbol{\beta}_{s},\boldsymbol{\delta}_{s},\boldsymbol{\theta}_{s},\boldsymbol{\varphi})
\end{equation}

To train the policy network $\boldsymbol{\varphi}$, the performance of classification accuracy with different labels is tested in different BER situations at the selected split points, and these results are used to constitute a new dataset to train the policy network. Based on this dataset, it is viable to get the optimal split point on the premise of ensuring that the semantic encoder does not exceed the device computing power, thus transforming the aforementioned dataset into a more comprehensive one with the optimal split point under the corresponding BER. The corresponding optimal split point can be obtained by optimizing the loss function:
\begin{equation}
\min_{\boldsymbol{\varphi}}\quad F_{\mathrm{MSE}}(\boldsymbol{h}, \hat{\boldsymbol{h}}),
\label{phi}
\end{equation}
where $F_{\mathrm{MSE}}\left(\boldsymbol{h}, \hat{\boldsymbol{h}} \right)  = - \sum_{j \in \Omega} \left\|\boldsymbol{h}[j]-\hat{\boldsymbol{h}}[j]\right\|^{2}$. $\boldsymbol{h}$ is the ground truth according to the optimal split point that we have tested, and $\hat{\boldsymbol{h}}$ denotes the configuration information that is deduced by the policy network. 

\subsection{The Whole Network Training}

\begin{algorithm}[t]
	\renewcommand{\algorithmicrequire}{\textbf{Initialization:}}
	\renewcommand{\algorithmicensure}{\textbf{Output:}}
	\caption{The Training Process of the Joint Device-Edge Digital Semantic Communication Network}
	\label{alg:1}
	\begin{algorithmic}[1]
		\REQUIRE  The ideal network $\mathbf{\Psi}, T=1$.
		\ENSURE $\boldsymbol{\alpha}_{s},\boldsymbol{\delta}_{s},\boldsymbol{\theta}_{s}, \boldsymbol{\beta}_{s}$
		\STATE {\bf Input}:The dateset $\mathcal{K}$
		\FOR{$s=1,\ldots,s_{\text{max}}$}
		\FOR{$Epo=1,\ldots,300$}
		\STATE Update $T$ based on Eq. (\ref{T})
		\STATE Train $\boldsymbol{\alpha}_{s},\boldsymbol{\beta}_{s},\boldsymbol{\theta}_{s}, \boldsymbol{\delta}_{s}$ with the loss Eq. (\ref{step1})
		\ENDFOR
		\ENDFOR
	\end{algorithmic}  
	\renewcommand{\algorithmicrequire}{\textbf{Initialization:}}
	\renewcommand{\algorithmicensure}{\textbf{Output:}}
	\begin{algorithmic}[1]
		\REQUIRE  The trained $\boldsymbol{\alpha}_{s},\boldsymbol{\delta}_{s},\boldsymbol{\theta}_{s},\boldsymbol{\beta}_{s}, T=1$.
		\ENSURE  $\boldsymbol{\alpha}_{s},\boldsymbol{\delta}_{s},\boldsymbol{\theta}_{s}, \boldsymbol{\beta}_{s}$
		\STATE {\bf Input}:The dateset $\mathcal{K}$
		\FOR{$s=1,\ldots,s_{\text{max}}$}
		\FOR{$Epo=1,\ldots,300$}
		\STATE Update $T$ based on Eq. (\ref{T})
		\STATE Train $\boldsymbol{\alpha}_{s},\boldsymbol{\beta}_{s},\boldsymbol{\theta}_{s}$, sparse $\boldsymbol{\delta}_{s}$ with the loss Eq. (\ref{step2})
		\ENDFOR
		\ENDFOR
	\end{algorithmic}  

	\renewcommand{\algorithmicrequire}{\textbf{Initialization:}}
	\renewcommand{\algorithmicensure}{\textbf{Output:}}
	\begin{algorithmic}[1]
		\REQUIRE The trained $\boldsymbol{\alpha}_{s},\boldsymbol{\delta}_{s},\boldsymbol{\theta}_{s},\boldsymbol{\beta}_{s},\boldsymbol{\varphi}, T \rightarrow \infty$.
		\ENSURE  $\boldsymbol{\varphi}$
		\STATE {\bf Input}:The dateset $\mathcal{K}$
		\STATE Freeze $\boldsymbol{\alpha}_{s},\boldsymbol{\delta}_{s},\boldsymbol{\theta}_{s}, \boldsymbol{\beta}_{s}$
		\FOR{$s=1,\ldots,s_{\text{max}}$}
		\FOR{$Epo=1,\ldots,300$}
                \STATE $\text{BER} = rand(10^{-1},10^{-5})$
			\STATE Train $\boldsymbol{\varphi}$ with the loss Eq. (\ref{phi})
		\ENDFOR
		\ENDFOR

	\end{algorithmic}
\end{algorithm}

The whole process of network training is illustrated in Algorithm \ref{alg:1}. In specific, the entire training process of this semantic communication system can be divided into three steps. The baseline is the ideal network with the noiseless channel. The wireless channel used in this experiment is BSC. 

Firstly, we jointly train the encoder, decoder, structured pruning and quantization module for $Epo=300$ epochs. The parameter $T$ in quantization is set to 1 and increased with the training epoch $Epo$, which can be given as:
\begin{equation}
\begin{aligned}
T = 
\left\{\begin{aligned}
&[Epo / 10]^{+},&Epo \leq 100, \\
&Epo,  &Epo > 100 ,
\end{aligned}\right.
\end{aligned}
\label{T}
\end{equation}
where $[\cdot]^{+}$ is the rounded up function. We use an invariant $\text{BER}_{\text{train}}=10^{-2}$ for training to achieve stable performance, while BERs in the range from $\text{BER}_{\text{test}}=10^{-1}$ to $10^{-5}$ are tested. This training process can be expressed as:
\begin{equation}
\min _{\boldsymbol{\alpha}_{s},\boldsymbol{\beta}_{s},\boldsymbol{\delta}_{s},\boldsymbol{\theta}_{s}} \mathcal{L}_{\text{se}}(\boldsymbol{\alpha}_{s},\boldsymbol{\beta}_{s},\boldsymbol{\delta}_{s},\boldsymbol{\theta}_{s}).
\label{step1}
\end{equation}

Then, we load the trained network of the previous process and jointly retrain the encoder, decoder, quantization and structured pruning for 300 epochs, by further taking into account the constraints of the structured pruning $\boldsymbol{\delta}_{s}$. Since the performance of convergence is depressed when using $l_{1}$-regularization to enforce the $\boldsymbol{\delta}_{s}$ to be sparse, we train this parameter separately. The experimental setup is the same as the previous process. This training process aims at:
\begin{equation}
\min _{\boldsymbol{\alpha}_{s},\boldsymbol{\beta}_{s},\boldsymbol{\delta}_{s},\boldsymbol{\theta}_{s}} \mathcal{L}_{\text{se}}(\boldsymbol{\alpha}_{s},\boldsymbol{\beta}_{s},\boldsymbol{\delta}_{s},\boldsymbol{\theta}_{s}) + \gamma \mathcal{L}_{\text{c}}(\boldsymbol{\delta}_{s}).
\label{step2}
\end{equation}

Finally, we load and freeze the trained encoder, decoder, structured pruning and quantizer. For the policy network, we vary channel BERs, from $\text{BER}_{\text{test}}=10^{-1}$ to $10^{-5}$ for training to search the optimal split points in the whole network by optimizing Eq. (\ref{phi}).

\section{numerical results}
In this section, we evaluate the performance of the proposed joint device-edge semantic communication architecture. We also evaluate the effectiveness of each component through ablation studies.

\subsection{Simulation Settings}
\begin{table}
\begin{center}
\caption{Setup for the digtial semantic communication system.}
\begin{tabular}{c|cc}
\hline
 & CIFAR-10 & ImageNet \\
\hline
batch size & 1024 & 256 \\  
\hline
image size & 32×32 & 256×256 \\  
\hline
learning rate & $10^{-3}$ & $10^{-3}$ \\  
\hline
optimizer & Adam & Adam \\  
\hline
\end{tabular}
\label{tab:hyperparameters}
\end{center}
\end{table}
We employ the CIFAR-10 image dataset for training \cite{2012Convolutional}. The dataset consists of 60000 RGB images divided into 10 different classes. Each class includes 6000 images of size 32 × 32 pixels, 5000 for training and 1000 for testing. We also train the digtial semantic communication system on the ImageNet \cite{5206848} dataset. We provide an overview of the experimental setup of the system in Table \ref{tab:hyperparameters}.

In the experiment, we decompose a complete ResNet56 \cite{Identity} to obtain the encoder and decoder networks. For ResNet, the residual block is a bottleneck block, which consists of 3 residual units and 1 residual connection. Each unit has a convolutional filter followed by ReLU and BN. The simulation is carried out by the computer with Intel Xeon Silver 4110 CPU @2.10GHz and NVIDIA RTX A4000.

\subsection{Performance of the Joint Device-Edge Semantic System}
\begin{figure}
\centering
\includegraphics[width=0.5\linewidth]{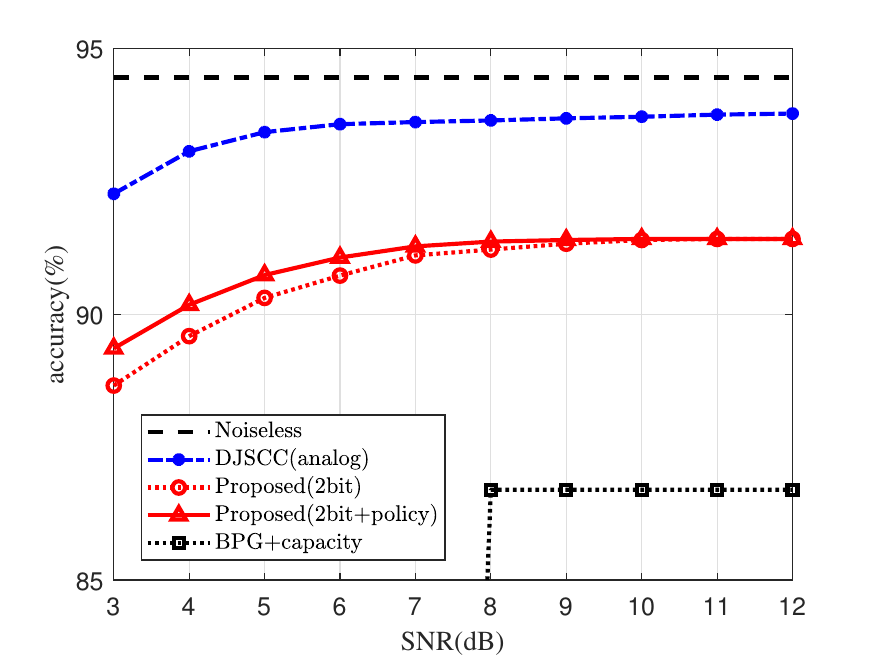}
\caption{Performance of the proposed digital semantic communication and the benchmarks on CIFAR-10 dataset.}
\label{fig:summary}
\end{figure}
\begin{figure}
\centering
\includegraphics[width=0.5\linewidth]{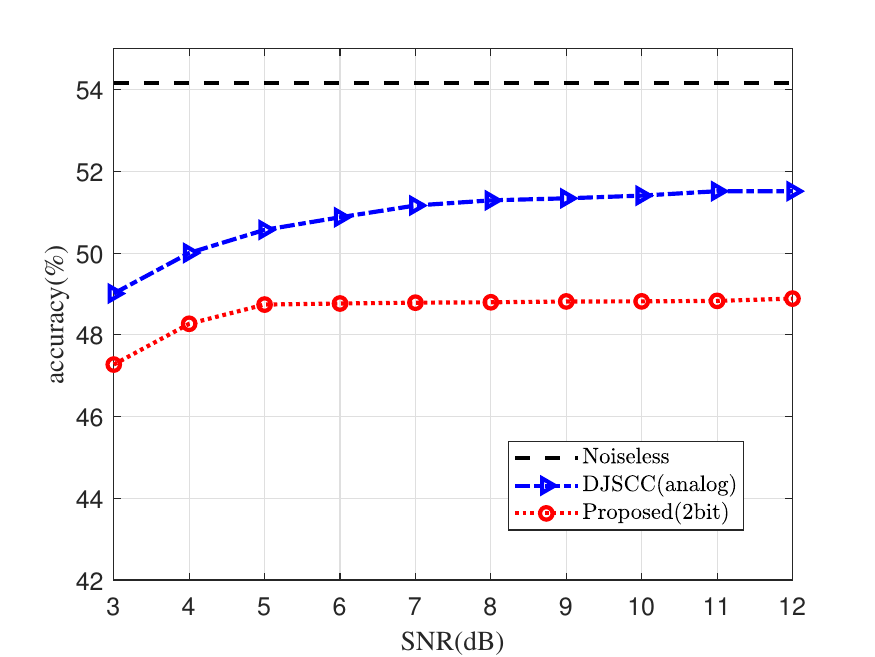}
\caption{Comparison between the analog semantic communication and the proposed digital semantic communication on the ImageNet \cite{5206848} dataset.}
\label{fig:summary_imagenet}
\end{figure}

To compare with the analog semantic communication, we assume quadrature phase shift keying (QPSK), and map the BERs of the BSC into corresponding SNRs of the additive white gaussian noise (AWGN) channel. 

The legend in Fig. \ref{fig:summary} with ``Proposed (2bit+policy)'' represents our digital semantic communication system with adaptive network split. In the scheme ``Proposed (2bit)'', we fix the split point of the encoding and decoding networks at the 9-th layer without using the policy network. The traditional method with source coding BPG and ideal capacity achieving channel coding is evaluated, which is labled as ``BPG+capacity'' in Fig. \ref{fig:summary}. While DJSCC uses arbitrary complex values for transmission in the analog form, the digital system uses a discrete set of quantified values as channel inputs. DJSCC gives the highest achievable performance of the digital semantic system \cite{9998051}, although it raises other challenges in the integration with existing communication systems. The results show that our proposed digital semantic communication system has significant performance improvement compared to the traditional method ``BPG+capacity'', and is close to the analog semantic communication system. When $\text{SNR}=6\text{dB}$, the accuracy loss caused by quantization is only 2\%. When the digital semantic communication system is combined with a policy network, its performance can be more close to that of the analog system, and better than that with the fixed split point.

We also compare the analog and the digtial semantic communication system on the ImageNet with a resolution of 256 × 256. Five epochs with the parameter $T=1$ of the quantization module are used to make the network with the quantization levels converge better. $T$ is then increased from 1 to 10 for the next five epochs. Finally, the system is trained for two more epochs with $T=100$. Fig. \ref{fig:summary_imagenet} illustrates the performance of the digital semantic communication system on the ImageNet dataset.

\subsection{Improvement of the Policy Network}

\begin{figure}
\centering
\includegraphics[width=0.5\linewidth]{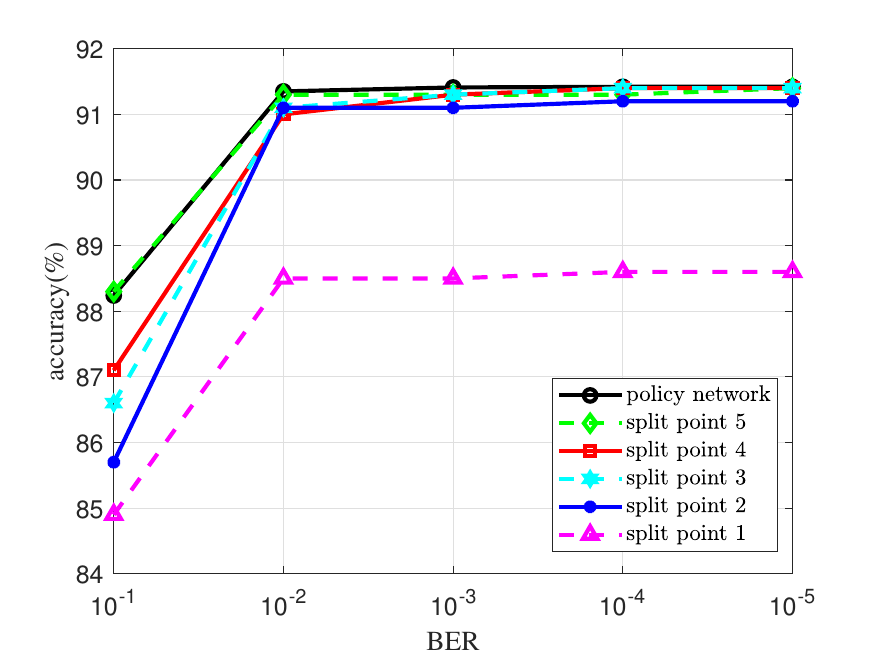}
\caption{Performance of the proposed digital semantic communication using the policy network and the fixed split point.}
\centering
\label{policy:a}
\end{figure}

\begin{figure}
\centering
\includegraphics[width=0.5\linewidth]{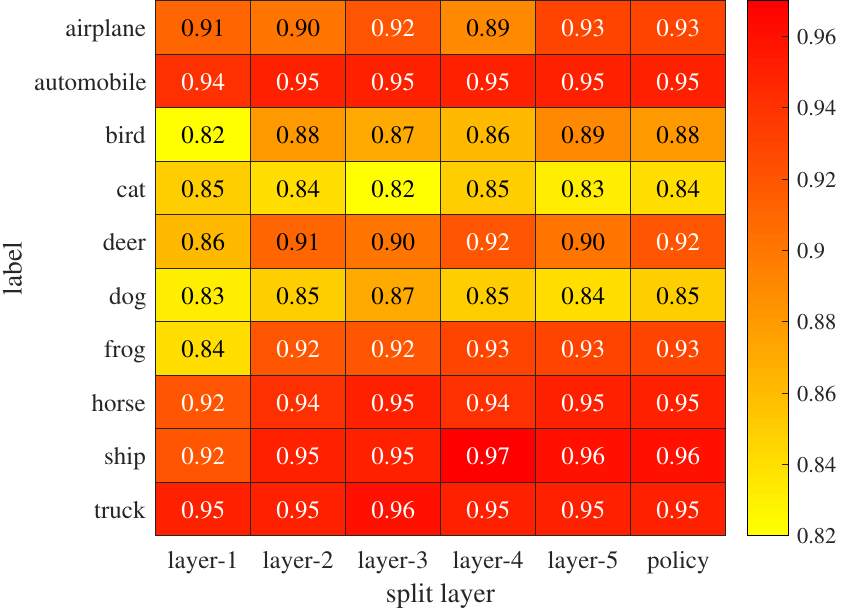}
\caption{Accuracy of different labels when using the policy network and the fixed split point under BER=0.01.}
\centering
\label{policy:b}
\end{figure}

We follow the training process outlined in Section III-E to train the adaptive digital semantic communication system. The results of this training process are illustrated in Fig. \ref{policy:a}. The experimental results show that the performance of the selection of the split point by the policy network approaches that of the ensemble of the networks with fixed split points when $\text{BER} \leq 0.01$ with negligible difference. The performance of the system with policy network is 0.9\% lower than that of the ensemble of system with the fixed split points when $\text{BER} \ge 0.01$. To more effectively compare the benefits brought by the policy network, we visualize the accuracy of different labels at different split points and under the guidance of the policy network when $\text{BER} = 0.01$, as shown in the Fig. \ref{policy:b}. Under the guidance of the policy network, the system almost achieve optimal or suboptimal performance for classifying different labels. This indicates that the policy network can choose more suitable split points given different images and channel conditions, thus ensuring an overall improvement in accuracy.

\subsection{Ablation Study}

\begin{table}
\begin{center}
\caption{Ablation study results for the different modules when BER=0.01.}
\begin{tabular}{c|ccccc|c|c}
\hline
 &\!pruning\!&\!quantization\!&\!SLL\!&CE&\!noise\!&\!ACC(\%)\!&\!$z$\! \\
\hline
case 1 & &  &  & \Checkmark  & & 94.46 & 128 \\  
\hline
case 2 &  &  &  & \Checkmark  & \Checkmark & 93.89 & 128 \\  
\hline
case 3 & &  & \Checkmark & & \Checkmark & 94.62 & 128 \\  
\hline
case 4 & \Checkmark & & & \Checkmark & \Checkmark & 93.16 & 13 \\  
\hline
case 5 &\Checkmark & & \Checkmark & & \Checkmark &93.58 & 13 \\  
\hline
case 6 & \Checkmark & \Checkmark & & \Checkmark & \Checkmark & 90.85 & 13\\  
\hline
case 7 &\Checkmark & \Checkmark & \Checkmark & &  \Checkmark & 91.14 & 13 \\  
\hline
\end{tabular}
\label{tab:Ablation}
\end{center}
\end{table}

We have incorporated structured pruning into the encoding network to remove redundancy in the semantic features and reduce the number of transmitted bits. To convert analog features into digital features, we have designed a non-linear quantization function with the trainable quantization level to replace the ideal quantization function. In the decoding network, we have proposed a novel SLL to better understand the semantic meaning distorted by noise. To verify the effectiveness of these methods, we conduct an ablation study with different combinations of these modules when $\text{BER}=0.01$. The results are shown in Table \ref{tab:Ablation}. Note that $z$ in Table \ref{tab:Ablation} represents the number of channels for the features, which is equivalent to the size of the transmitted data and also implies the bandwidth required for transmission. 

In this experiment, the setup is simplified by fixing the split point after the 9-th layer. The model only uses the CE loss function (case 1), which is also the model with the ideal channel in Fig. 7. The proposed model with SLL in the noise channel (case 3) outperforms the model with CE in the noiseless channel (case 1) in an analog fashion. The structured pruning module can reduce redundancy in semantic features, and results in a slight accuracy loss of about 1\% in analog semantic communication. This is evident when comparing cases 2 and 4 or cases 3 and 6. However, the bandwidth can be compressed to approximately $1/10$ of its original size ($z=128$ vs. $z=13$), leading to increased efficiency and decreased demand for communication resources in resource-constrained joint device-edge communication scenarios. The quantization module can be used to convert analog semantic communication into digital semantic communication, resulting in about 2\% performance loss when comparing cases 4 and 6 or cases 5 and 7. Furthermore, when the SLL is implemented in digital semantic communication, it achieves a higher accuracy to reduce the impact of the quantization module and pruning module when comparing cases 4 to 5 and cases 6 to 7. The results demonstrate that the structured pruning and SLL can significantly improve communication efficiency and model performance for joint device-edge digital semantic communication systems.

\section{Conclusion}
In this paper, we have proposed a joint device-edge digital semantic communication system. The system comprises an encoding network deployed on the resource-limited device and a decoding network deployed at the edge. To minimize the bandwidth cost of transmitting the semantic features, we have incorporated structured pruning and quantization network into the encoding network. We have adopted a sparse scaling factor to eliminate redundant features and reduce the required bandwidth. To achieve better semantic representation for digital transmission, we have proposed a novel non-linear quantization method with trainable quantization levels. Additionally, we have introduced an SLL function to reduce semantic error, which is used to learn ``soft'' semantic information to enhance comprehension in the presence of channel noise. To enhance the flexibility and reliability of the semantic communication network, we have proposed a policy network that dynamically determine the optimal configuration of the encoding network, by maximizing the classification accuracy under the constraint of the limited computing resources. The policy network adaptively selects the split point and transmitted semantic features based on diverse channel conditions and image properties. Our experimental results have demonstrated that the proposed semantic communication network achieves satisfactory accuracy compared to DJSCC in the classification task on the CIFAR-10 dataset and ImageNet dataset. Furthermore, the ablation study has demonstrated the effectiveness of the proposed modules, including the quantization module, structured pruning and SLL.

\bibliographystyle{IEEEtran}
\bibliography{IEEEabrv,baiduxueshu_papers}
\end{document}